\documentclass[a4paper,aps,prd,10pt,preprintnumbers,showpacs,twocolumn,superscriptaddress,nofootinbib,amsmath,amssymb,floatfix]{revtex4-1}
\usepackage{graphicx}
\usepackage{cmap}
\usepackage[utf8]{inputenc}
\usepackage[T1]{fontenc}
\def\imo{i}

\def\im#1{Im(#1)}
\def\K{{\cal K}}
\def\Order#1{{\cal O}\left(#1\right)}

\newcommand{\inlinefigurecaption}[2]{%
  \refstepcounter{figure}%
  \label{#2}%
  \begingroup\small\textbf{FIG.~\thefigure.} #1\par\endgroup
}
\newcommand{\inlinetablecaption}[1]{%
  \refstepcounter{table}%
  \begingroup\small\textbf{TABLE~\thetable.} #1\par\endgroup
}

\begin{document}
\title{Quasinormal modes and excitation factors of a regular black hole with zero-point length}
\author{Milena Skvortsova}\email{milenas577@mail.ru}
\affiliation{Peoples' Friendship University of Russia (RUDN University), 6 Miklukho-Maklaya Street, Moscow, 117198, Russia}
\begin{abstract}
We study the ringdown of the regular Jusufi-Singleton black hole, whose nonsingular core is controlled by a zero-point length arising from a non-local, T-duality-inspired gravitational model.  Scalar, electromagnetic and Dirac perturbations are considered.  The zero-point-length parameter raises the effective scattering barrier and produces a systematic increase of the oscillation frequencies, while also making the damping faster over most of the parameter range.  High-order WKB results are checked against time-domain integration and show very good agreement for the dominant modes.  We also compute excitation factors, which characterize the source-independent strength of the quasinormal-mode poles and show a smooth dependence on the new length scale.
\end{abstract}
\maketitle
\section{Introduction}

Quasinormal modes are one of the most useful probes of black-hole spacetimes \cite{Kokkotas:1999bd,Berti:2009kk,Konoplya:2011qq,Nollert:1999ji,Bolokhov:2025rng}.  They determine the ringdown stage after a perturbation and are fixed by the geometry and by the spin of the perturbing field, rather than by the details of the initial disturbance.  For this reason they provide a clean way to compare different compact-object backgrounds and to identify how modifications of the near-horizon \cite{Konoplya:2022pbc} or photon-orbit geometry \cite{Cardoso:2008bp} are reflected in observable ringing frequencies.

Regular black holes are especially interesting in this context.  They replace the central curvature singularity by a finite core while preserving an event horizon, and therefore offer a useful phenomenological arena for testing possible short-distance modifications of gravity.  Many such metrics are supported by nonlinear electrodynamics, effective quantum-gravity corrections, or other modified sources, and their quasinormal spectra have been studied in a variety of models~\cite{Hayward:2005gi, Spina:2025wxb,Ayon-Beato:1998hmi,Dymnikova:1992ux,Konoplya:2024kih,Bronnikov:2000vy,Konoplya:2025ect,Nicolini:2005vd,Bonanno:2025dry,Bronnikov:2024izh,Bolokhov:2024sdy}. In these geometries the ringdown, scattering and Hawking radiation is sensitive not only to the position of the horizon, but also to how the effective potential is deformed outside the black hole \cite{Guo:2024jhg,Konoplya:2023aph,Skvortsova:2025cah,Li:2014fka,Bolokhov:2023ruj,Lopez:2022uie,Panotopoulos:2019qjk,Skvortsova:2024wly,Jawad:2020hju,Flachi:2012nv,Held:2019xde,Mukohyama:2023xyf,Konoplya:2023ppx,Meng:2022oxg,Dubinsky:2026wcv,Lin:2013ofa,Yang:2021cvh,Macedo:2016yyo,Pedraza:2021hzw,Gingrich:2024tuf,Huang:2023aet,Bolokhov:2025fto,Al-Badawi:2023lke,Konoplya:2025hgp,Cai:2021ele,DuttaRoy:2022ytr,Jusufi:2020odz,Skvortsova:2026unq,MahdavianYekta:2019pol,Konoplya:2023aph,Zhang:2024nny,Bolokhov:2026eqf,Arbey:2021jif,Arbey:2021yke,Arbey:2026koc,Vagnozzi:2022moj,Calza:2024fzo,Calza:2024xdh,Calza:2025mwn,Pedrotti:2024znu,Calza:2024qxn,Calza:2025yfm}.

The spacetime considered in the present work is the neutral regular black hole proposed by Jusufi and Singleton~\cite{Jusufi:2025selfenergy}.  Its origin is different from the standard charged regular black holes: the regularizing parameter is a zero-point length $l_0$ associated with a non-local, T-duality-inspired description of the source.  The same scale regularizes the gravitational self-energy and allows this finite self-energy to be included in the effective mass distribution.  Thus $l_0$ controls both the size of the regular core and the deviation of the exterior potential from the Schwarzschild-like case.  This makes the solution a natural background for asking how a non-local short-distance scale affects black-hole spectroscopy. While grey-body factors and Hawking radiation of these regular black holes have been considered recently in \cite{3164933}, no such analysis is known for quasinormal frequencies.

Our goal is to characterize the ringdown around such regular black holes.   We compute quasinormal frequencies of fields of various spin using high-order WKB expansions with Pad\'e resummation and check representative modes by direct time-domain integration followed by Prony extraction.  In addition, we calculate excitation factors, namely the source-independent residues of the frequency-domain Green function at the quasinormal poles.  The latter contain information that is complementary to the frequencies: they measure the intrinsic strength of a pole in a fixed scattering normalization, while the actual waveform amplitude still depends on the source or initial data.

The paper is organized as follows.  In Sec.~\ref{sec:wavelike} we summarize the Jusufi-Singleton geometry and the effective potentials for scalar, electromagnetic and Dirac perturbations.  Section~\ref{sec:methods} describes the WKB--Pad\'e and time-domain methods.  The quasinormal spectra and their dependence on the zero-point length are discussed in Sec.~\ref{sec:qnm}.  Section~\ref{sec:excitation} gives the excitation-factor calculation and the corresponding numerical results.  Conclusions are presented in the final section.

\section{Geometry and perturbation equations}\label{sec:wavelike}

The background considered here is the neutral black-hole spacetime obtained in Ref.~\cite{Jusufi:2025selfenergy} from a non-local gravitational theory inspired by T-duality.  In this construction a point source is replaced by an extended distribution characterized by a zero-point length $l_0$.  The same non-local scale regularizes the Newtonian self-energy of the gravitational field and permits one to include this finite self-energy as an effective contribution to the spacetime source.  As a result, the ADM mass receives a regularized gravitational contribution, producing a nonsingular, electrically neutral geometry of Ay\'on-Beato--Garc\'ia type.  In contrast with charged regular black holes, the regularizing scale here is not an electric charge but the zero-point length of the non-local theory.  The line element can be written in the standard static and spherically symmetric form
\begin{equation}\label{metric}
  ds^2=-f(r)dt^2+\frac{dr^2}{f(r)}+r^2(d\theta^2+\sin^2\theta d\phi^2).
\end{equation}
For the Jusufi-Singleton solution used in this work, the metric function is
\begin{widetext}
\begin{eqnarray}
 f(r)&=&1-\frac{2 M r^2}{\left(l_0^2+r^2\right)^{3/2}}
 -\frac{3M^2}{8l_0 r}\tan^{-1}\left(\frac{r}{l_0}\right)
 +\frac{3l_0^2M^2r^2}{8\left(l_0^2+r^2\right)^2}
 +\frac{5M^2r^2}{8\left(l_0^2+r^2\right)^2} .
\end{eqnarray}
\end{widetext}
Here $M$ denotes the mass parameter and $l_0$ controls the size of the non-local core.  At large radii the metric approaches the Schwarzschild form with corrections generated by the finite self-energy term, while near the center the even dependence on $r$ makes the geometry smooth.  Throughout the numerical calculations below we measure all dimensional quantities in units of $M$ and set $M=1$.  The parameter $l_0$ then measures the strength of the zero-point-length correction.  For the values used in the figures and tables the metric has an event horizon and the exterior problem is a standard black-hole scattering problem.

We treat scalar, electromagnetic and Dirac fields as test perturbations of this fixed geometry, neglecting their backreaction on the spacetime.  The corresponding covariant field equations are
\begin{subequations}\label{coveqs}
\begin{align}
\frac{1}{\sqrt{-g}}\partial_\mu
\left(\sqrt{-g}g^{\mu\nu}\partial_\nu\Phi\right)&=0,
\label{KGg}\\
\nabla_\mu F^{\mu\nu}&=0,
\label{EmagEq}\\[-1mm]
F_{\mu\nu}&=\partial_\mu A_\nu-\partial_\nu A_\mu,
\label{EmagTensor}\\
\gamma^{\alpha}\left(\frac{\partial}{\partial x^{\alpha}}
-\Gamma_{\alpha}\right)\Upsilon&=0.
\label{covdirac}
\end{align}
\end{subequations}
where $\Gamma_{\alpha}$ is the spin connection in the tetrad formalism.  After separation into spherical harmonics and a harmonic time dependence $e^{-i\omega t}$, each perturbation reduces to a one-dimensional wave equation of the Schr\"odinger type,
\begin{equation}\label{wave-equation}
\dfrac{d^2 \Psi}{dr_*^2}+\left(\omega^2-V(r)\right)\Psi=0,
\end{equation}
where the tortoise coordinate is defined by
\begin{equation}\label{tortoise}
\frac{dr_*}{dr}=\frac{1}{f(r)}.
\end{equation}
Thus the horizon is mapped to $r_*\to -\infty$, while spatial infinity corresponds to $r_*\to +\infty$.

For scalar ($s=0$) and electromagnetic ($s=1$) perturbations the effective potential is (see, for example, \cite{Carter:1968ks,Konoplya:2018arm})
\begin{eqnarray}\label{potentialScalar}
V_s(r)&=&f(r)\frac{\ell(\ell+1)}{r^2}+\frac{1-s}{r}\frac{d^2 r}{dr_*^2}
\nonumber\\
&=&f(r)\frac{\ell(\ell+1)}{r^2}+(1-s)\frac{f(r)f'(r)}{r},
\end{eqnarray}
with $\ell=s,s+1,s+2,\ldots$.  The second term is present for scalar perturbations and absent for electromagnetic perturbations.  For the massless Dirac field one obtains the pair of supersymmetric partner potentials
\begin{equation}
V_{\pm}(r)=W^2\pm\frac{dW}{dr_*}, \qquad
W=\left(\ell+\frac{1}{2}\right)\frac{\sqrt{f(r)}}{r},
\end{equation}
where $\ell=1/2,3/2,\ldots$.  The two Dirac wave functions are related by the Darboux transformation (see \cite{Cho:2003qe,Konoplya:2007zx,Konoplya:2017tvu,Cho:2004wj,Kanti:2006ua} and references therein)
\begin{equation}\label{psi}
\Psi_{+}\propto \left(W+\dfrac{d}{dr_*}\right) \Psi_{-},
\end{equation}
so that $V_+$ and $V_-$ are isospectral.  We therefore use $V_+$ in the numerical analysis.

\begin{center}
\resizebox{\linewidth}{!}{\includegraphics{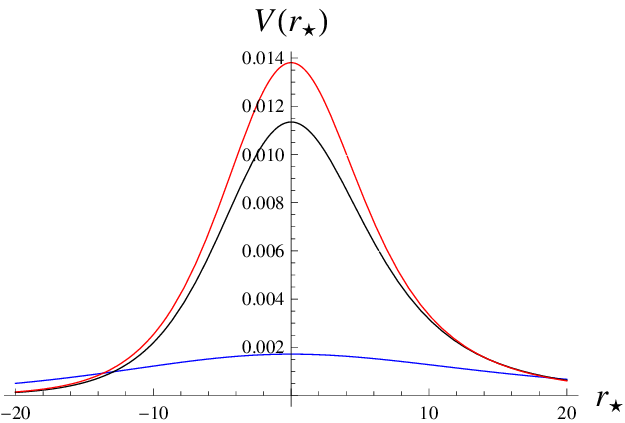}}
\inlinefigurecaption{Scalar effective potential $V(r_*)$ for the $\ell=0$ mode of the Jusufi-Singleton black hole with $M=1$.  The curves correspond to $l_0=0.1$ (blue), $l_0=0.5$ (black) and $l_0=0.65$ (red).  Increasing $l_0$ raises the barrier and makes the peak more pronounced, which is reflected in the larger oscillation frequencies in the WKB spectrum.}{fig:pot1}
\end{center}

\begin{center}
\resizebox{\linewidth}{!}{\includegraphics{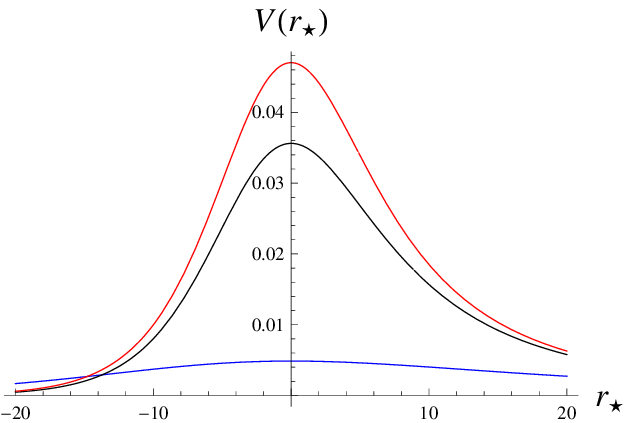}}
\inlinefigurecaption{Electromagnetic effective potential $V(r_*)$ for the $\ell=1$ mode at $M=1$.  The color coding is the same as in Fig.~\ref{fig:pot1}.  The potential remains a single positive barrier for all three values of $l_0$, and the barrier height grows as the zero-point length is increased.}{fig:pot2}
\end{center}

\begin{center}
\resizebox{\linewidth}{!}{\includegraphics{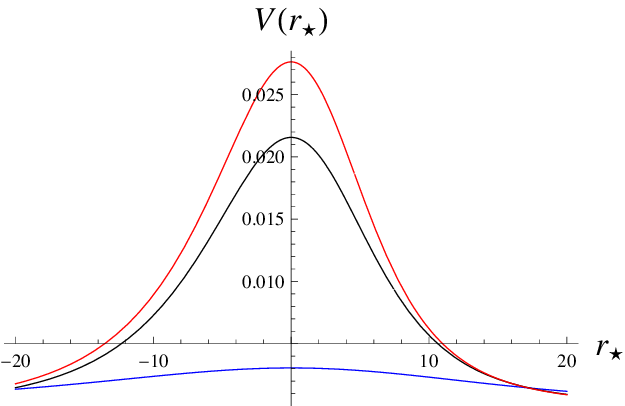}}
\inlinefigurecaption{Dirac effective potential $V_+(r_*)$ for the $\ell=1/2$ mode at $M=1$.  The curves correspond to $l_0=0.1$ (blue), $l_0=0.5$ (black) and $l_0=0.65$ (red).  The Darboux-related potential $V_-$ gives the same quasinormal spectrum.  As in the scalar and electromagnetic cases, the dominant effect of increasing $l_0$ is to increase the height and curvature of the scattering barrier near its maximum.}{fig:pot3}
\end{center}

Figures~\ref{fig:pot1}--\ref{fig:pot3} show that the perturbation problem has the expected black-hole barrier structure: the potentials vanish toward the horizon and at spatial infinity, while a single peak controls the ringdown spectrum.  This behavior justifies the use of WKB methods for the low-lying modes considered below.  The scalar monopole in Fig.~\ref{fig:pot1} has the lowest barrier, so it produces the smallest oscillation frequency.  The electromagnetic dipole in Fig.~\ref{fig:pot2} is higher because the centrifugal contribution is not offset by the scalar derivative term.  The Dirac potential in Fig.~\ref{fig:pot3} has the same qualitative single-peak form and is representative of the isospectral pair $V_\pm$.  In all three cases, increasing $l_0$ raises the peak and changes its curvature near the maximum; this explains the monotonic growth of the real part of the fundamental quasinormal frequencies displayed in the tables.

\section{Methods for quasinormal modes}\label{sec:methods}

We compute the quasinormal frequencies by using two complementary procedures.  The main data in the tables are obtained from a high-order WKB expansion with Pad\'e resummation, which is well suited to the single-barrier potentials shown in Figs.~\ref{fig:pot1}--\ref{fig:pot3}.  As an independent check, we also evolve the perturbation equation in the time domain and extract the ringing frequencies from the resulting signal by the Prony method.

\subsection{Higher-order WKB method}\label{subsec:wkb}

For each spin sector the radial problem is reduced to Eq.~(\ref{wave-equation}).  Quasinormal modes are selected by imposing wave behavior with no incoming radiation from infinity and no outgoing radiation from the horizon,
\begin{equation}\label{boundaryconditions}
\Psi(r_*\to\pm\infty)\propto e^{\pm\imo \omega r_*},
\end{equation}
where the minus sign corresponds to a purely ingoing wave at $r_*\to-\infty$, and the plus sign to a purely outgoing wave at $r_*\to\infty$.  The WKB construction uses the Taylor series of the effective potential at its maximum and matches the local solution near this maximum to the asymptotic waves satisfying Eq.~(\ref{boundaryconditions}).  At leading order it reproduces the eikonal approximation, while higher orders incorporate successive derivatives of the potential at the peak.  In the notation of Ref.~\cite{Konoplya:2019hlu}, the expansion can be written as
\begin{eqnarray}\label{WKBformula-spherical}
\omega^2&=&V_0+A_2(\K^2)+A_4(\K^2)+A_6(\K^2)+\ldots\\\nonumber&-&\imo \K\sqrt{-2V_2}\left(1+A_3(\K^2)+A_5(\K^2)+A_7(\K^2)\ldots\right),
\end{eqnarray}
with the usual quantization condition
\begin{equation}
\K=n+\frac{1}{2}, \quad n=0,1,2,\ldots,
\end{equation}
where $n$ denotes the overtone number.  The quantities $V_0$ and $V_2$ are, respectively, the value of the potential and its second derivative with respect to $r_*$ at the maximum.  The correction $A_i$ depends on $\K$ and on derivatives of the potential at the peak up to order $2i$.  The terms of second and third WKB order were derived in Ref.~\cite{Iyer:1986np}, the fourth--sixth orders in Refs.~\cite{Konoplya:2003ii}, and the seventh--nineteenth orders in Ref.~\cite{Matyjasek:2017psv,Matyjasek:2019eeu,Matyjasek:2026yiu}. In the numerical implementation below we use the WKB expansion at higher orders and with Pad\'e approximants and compare the 16th-order and 14th-order resummations as an internal estimate of the stability of the result.  The WKB method at various orders has been applied to quasinormal spectra and grey-body factors in many contexts
(see, for example, \cite{Kokkotas:2010zd,Dubinsky:2024rvf,Dubinsky:2024hmn,Konoplya:2010vz,Guo:2020blq,Konoplya:2007yy,Tan:2022vfe,Albuquerque:2023lhm,Malik:2023bxc,Fernando:2016ftj,Kodama:2009bf,Konoplya:2009hv,Dubinsky:2025wns,Dubinsky:2025ypj,Malik:2026jzl,Malik:2024tuf,Lutfuoglu:2026zxj,Lutfuoglu:2026zel,Konoplya:2023moy,Lutfuoglu:2025kqp,Konoplya:2019ppy}. The corresponding Pad\'e approximant  has the form \cite{Konoplya:2019hlu},
\begin{equation}
P_{\tilde m/\tilde n}(\varepsilon)=
\frac{\sum_{i=0}^{\tilde m} a_i \varepsilon^i}
     {\sum_{j=0}^{\tilde n} b_j \varepsilon^j},
\end{equation}
where $\tilde m$ and $\tilde n$ denote the orders of the numerator and denominator polynomials, respectively, while $\tilde m+\tilde n$ coincides with the order of the underlying WKB expansion. In most applications, the Pad\'e-resummed expression provides a substantially more accurate approximation than the corresponding truncated WKB series. Following recent studies \cite{Skvortsova:2026idf,Skvortsova:2026jtx,Lutfuoglu:2026rqe,Lutfuoglu:2026boa,Bolokhov:2026dfg,Bolokhov:2026uol}, we adopt symmetric Pad\'e approximants with $\tilde m=\tilde n$ for the 14th- and 16th-order WKB calculations, as this choice has consistently been found to yield the highest accuracy.

\subsection{Time-domain integration and Prony analysis}\label{subsec:timedomain}

To check the WKB--Pad\'e frequencies with a method that does not rely on an expansion at the potential maximum, we also integrate Eq.~(\ref{wave-equation}) directly in the time domain.  Introducing the null coordinates $u=t-r_*$ and $v=t+r_*$, the wave equation can be evolved on a characteristic grid.  We use the Gundlach-Price-Pullin finite-difference prescription \cite{Gundlach:1993tp},
\begin{eqnarray}
\Psi\left(N\right)&=&\Psi\left(W\right)+\Psi\left(E\right)-\Psi\left(S\right)\nonumber\\
&&- \Delta^2V\left(S\right)\frac{\Psi\left(W\right)+\Psi\left(E\right)}{8}+{\cal O}\left(\Delta^4\right),\label{Discretization}
\end{eqnarray}
where $N\equiv(u+\Delta,v+\Delta)$, $W\equiv(u+\Delta,v)$, $E\equiv(u,v+\Delta)$, and $S\equiv(u,v)$ are the four vertices of an elementary null-grid cell.  This characteristic scheme has been used extensively for black-hole perturbations \cite{Konoplya:2014lha,Konoplya:2005et,Dubinsky:2025nxv,Lutfuoglu:2026fpx,Dubinsky:2025bvf,Konoplya:2018yrp,Dubinsky:2024gwo,Abdalla:2012si,Konoplya:2023fmh,Bolokhov:2024ixe,Bolokhov:2024bke,Aneesh:2018hlp,Momennia:2022tug,Konoplya:2006gq,Konoplya:2013sba,Dubinsky:2024jqi,Malik:2024bmp,Malik:2024iky,Lutfuoglu:2025pzi,Skvortsova:2023zca}.

After the initial transient has passed, the waveform is dominated by damped ringing.  In this interval we fit the signal by a finite sum of exponents,
\begin{equation}
\Psi(t)\simeq \sum_{j=1}^{p} C_j e^{-\imo\omega_j t},
\end{equation}
which is the standard Prony representation of the ringdown stage.  The complex frequencies $\omega_j$ obtained from this fit are then compared with the WKB--Pad\'e values.  The dominant mode is selected from the part of the fit that remains stable under changes of the fitting window and of the number of exponents $p$.

\section{Quasinormal modes}\label{sec:qnm}

The quasinormal frequencies obtained by the WKB--Pad\'e method are collected in the following tables for scalar, electromagnetic and Dirac perturbations.  The time-domain profiles shown afterwards provide an independent check of the fundamental modes at the largest value of the non-local parameter, close to the extreme value, used in the numerical analysis.

\begin{center}
\inlinetablecaption{Fundamental scalar quasinormal modes of the Jusufi-Singleton black hole for $M=1$.  The mode label gives $(\ell,n)$, and all entries in this table have $n=0$.  The columns compare the 16th-order WKB result with Pad\'e approximant $\tilde{m}=8$ against the 14th-order WKB result with Pad\'e approximant $\tilde{m}=7$; $\Delta$ is the relative difference in percent.}
\begingroup
\scriptsize
\begin{ruledtabular}
\begin{tabular}{c c c c c}
Mode & $l_0$ & WKB16 ($\tilde{m}=8$) & WKB14 ($\tilde{m}=7$) & $\Delta$ \\
$(0,0)$ & $0.05$ & $0.016093-0.015245 i$ & $0.016081-0.015195 i$ & $0.230\%$\\
$(0,0)$ & $0.1$ & $0.028328-0.026669 i$ & $0.028305-0.026582 i$ & $0.230\%$\\
$(0,0)$ & $0.15$ & $0.038071-0.035564 i$ & $0.038038-0.035449 i$ & $0.230\%$\\
$(0,0)$ & $0.2$ & $0.046094-0.042683 i$ & $0.046052-0.042545 i$ & $0.230\%$\\
$(0,0)$ & $0.25$ & $0.052883-0.048494 i$ & $0.052830-0.048338 i$ & $0.229\%$\\
$(0,0)$ & $0.3$ & $0.058764-0.053303 i$ & $0.058700-0.053133 i$ & $0.229\%$\\
$(0,0)$ & $0.35$ & $0.063967-0.057314 i$ & $0.063894-0.057132 i$ & $0.228\%$\\
$(0,0)$ & $0.4$ & $0.068663-0.060664 i$ & $0.068583-0.060470 i$ & $0.228\%$\\
$(0,0)$ & $0.45$ & $0.072984-0.063445 i$ & $0.072901-0.063233 i$ & $0.235\%$\\
$(0,0)$ & $0.5$ & $0.077044-0.065711 i$ & $0.076945-0.065479 i$ & $0.249\%$\\
$(0,0)$ & $0.55$ & $0.080914-0.067480 i$ & $0.080787-0.067259 i$ & $0.241\%$\\
$(0,0)$ & $0.6$ & $0.084584-0.068715 i$ & $0.084497-0.068512 i$ & $0.202\%$\\
$(0,0)$ & $0.65$ & $0.088221-0.069215 i$ & $0.088167-0.069279 i$ & $0.0745\%$\\
$(1,0)$ & $0.05$ & $0.042666-0.014189 i$ & $0.042666-0.014189 i$ & $0\%$\\
$(1,0)$ & $0.1$ & $0.075080-0.024835 i$ & $0.075080-0.024835 i$ & $0\%$\\
$(1,0)$ & $0.15$ & $0.100864-0.033141 i$ & $0.100864-0.033141 i$ & $0\%$\\
$(1,0)$ & $0.2$ & $0.122068-0.039804 i$ & $0.122068-0.039804 i$ & $0\%$\\
$(1,0)$ & $0.25$ & $0.139978-0.045261 i$ & $0.139978-0.045261 i$ & $0\%$\\
$(1,0)$ & $0.3$ & $0.155458-0.049794 i$ & $0.155458-0.049794 i$ & $0\%$\\
$(1,0)$ & $0.35$ & $0.169125-0.053596 i$ & $0.169125-0.053596 i$ & $0\%$\\
$(1,0)$ & $0.4$ & $0.181441-0.056796 i$ & $0.181441-0.056796 i$ & $0\%$\\
$(1,0)$ & $0.45$ & $0.192769-0.059481 i$ & $0.192769-0.059481 i$ & $0\%$\\
$(1,0)$ & $0.5$ & $0.203410-0.061704 i$ & $0.203410-0.061704 i$ & $0\%$\\
$(1,0)$ & $0.55$ & $0.213623-0.063489 i$ & $0.213623-0.063489 i$ & $0\%$\\
$(1,0)$ & $0.6$ & $0.223652-0.064826 i$ & $0.223652-0.064826 i$ & $0\%$\\
$(1,0)$ & $0.65$ & $0.233741-0.065665 i$ & $0.233741-0.065665 i$ & $0\%$\\
$(2,0)$ & $0.05$ & $0.070440-0.014058 i$ & $0.070440-0.014058 i$ & $0\%$\\
$(2,0)$ & $0.1$ & $0.123948-0.024609 i$ & $0.123948-0.024609 i$ & $0\%$\\
$(2,0)$ & $0.15$ & $0.166502-0.032843 i$ & $0.166502-0.032843 i$ & $0\%$\\
$(2,0)$ & $0.2$ & $0.201489-0.039452 i$ & $0.201489-0.039452 i$ & $0\%$\\
$(2,0)$ & $0.25$ & $0.231030-0.044866 i$ & $0.231030-0.044866 i$ & $0\%$\\
$(2,0)$ & $0.3$ & $0.256555-0.049367 i$ & $0.256555-0.049367 i$ & $0\%$\\
$(2,0)$ & $0.35$ & $0.279080-0.053146 i$ & $0.279080-0.053146 i$ & $0\%$\\
$(2,0)$ & $0.4$ & $0.299371-0.056330 i$ & $0.299371-0.056330 i$ & $0\%$\\
$(2,0)$ & $0.45$ & $0.318027-0.059007 i$ & $0.318027-0.059007 i$ & $0\%$\\
$(2,0)$ & $0.5$ & $0.335546-0.061229 i$ & $0.335546-0.061229 i$ & $0\%$\\
$(2,0)$ & $0.55$ & $0.352364-0.063019 i$ & $0.352364-0.063019 i$ & $0\%$\\
$(2,0)$ & $0.6$ & $0.368896-0.064370 i$ & $0.368896-0.064370 i$ & $0\%$\\
$(2,0)$ & $0.65$ & $0.385566-0.065229 i$ & $0.385566-0.065229 i$ & $0\%$\\
\end{tabular}
\end{ruledtabular}
\endgroup
\end{center}

\begin{center}
\inlinetablecaption{Fundamental electromagnetic quasinormal modes of the Jusufi-Singleton black hole for $M=1$.  The mode label gives $(\ell,n)$ and separates the two fundamental multipoles shown here.  The last column gives the relative difference between the WKB16--Pad\'e and WKB14--Pad\'e approximants in percent.}
\begingroup
\scriptsize
\begin{ruledtabular}
\begin{tabular}{c c c c c}
Mode & $l_0$ & WKB16 ($\tilde{m}=8$) & WKB14 ($\tilde{m}=7$) & $\Delta$ \\
$(1,0)$ & $0.05$ & $0.036176-0.013441 i$ & $0.036176-0.013441 i$ & $0\%$\\
$(1,0)$ & $0.1$ & $0.063722-0.023543 i$ & $0.063722-0.023543 i$ & $0\%$\\
$(1,0)$ & $0.15$ & $0.085709-0.031441 i$ & $0.085709-0.031441 i$ & $0\%$\\
$(1,0)$ & $0.2$ & $0.103868-0.037796 i$ & $0.103868-0.037796 i$ & $0\%$\\
$(1,0)$ & $0.25$ & $0.119283-0.043016 i$ & $0.119283-0.043016 i$ & $0\%$\\
$(1,0)$ & $0.3$ & $0.132687-0.047370 i$ & $0.132687-0.047370 i$ & $0\%$\\
$(1,0)$ & $0.35$ & $0.144610-0.051038 i$ & $0.144610-0.051038 i$ & $0\%$\\
$(1,0)$ & $0.4$ & $0.155452-0.054144 i$ & $0.155452-0.054144 i$ & $0\%$\\
$(1,0)$ & $0.45$ & $0.165534-0.056768 i$ & $0.165534-0.056768 i$ & $0\%$\\
$(1,0)$ & $0.5$ & $0.175130-0.058959 i$ & $0.175130-0.058960 i$ & $0\%$\\
$(1,0)$ & $0.55$ & $0.184486-0.060739 i$ & $0.184486-0.060739 i$ & $0\%$\\
$(1,0)$ & $0.6$ & $0.193844-0.062091 i$ & $0.193844-0.062091 i$ & $0\%$\\
$(1,0)$ & $0.65$ & $0.203462-0.062956 i$ & $0.203462-0.062956 i$ & $0\%$\\
$(2,0)$ & $0.05$ & $0.066655-0.013805 i$ & $0.066655-0.013805 i$ & $0\%$\\
$(2,0)$ & $0.1$ & $0.117324-0.024170 i$ & $0.117324-0.024170 i$ & $0\%$\\
$(2,0)$ & $0.15$ & $0.157661-0.032264 i$ & $0.157661-0.032264 i$ & $0\%$\\
$(2,0)$ & $0.2$ & $0.190868-0.038766 i$ & $0.190868-0.038766 i$ & $0\%$\\
$(2,0)$ & $0.25$ & $0.218949-0.044098 i$ & $0.218949-0.044098 i$ & $0\%$\\
$(2,0)$ & $0.3$ & $0.243256-0.048536 i$ & $0.243256-0.048536 i$ & $0\%$\\
$(2,0)$ & $0.35$ & $0.264755-0.052266 i$ & $0.264755-0.052266 i$ & $0\%$\\
$(2,0)$ & $0.4$ & $0.284174-0.055413 i$ & $0.284174-0.055413 i$ & $0\%$\\
$(2,0)$ & $0.45$ & $0.302088-0.058064 i$ & $0.302088-0.058064 i$ & $0\%$\\
$(2,0)$ & $0.5$ & $0.318979-0.060269 i$ & $0.318979-0.060269 i$ & $0\%$\\
$(2,0)$ & $0.55$ & $0.335273-0.062051 i$ & $0.335273-0.062051 i$ & $0\%$\\
$(2,0)$ & $0.6$ & $0.351382-0.063398 i$ & $0.351382-0.063398 i$ & $0\%$\\
$(2,0)$ & $0.65$ & $0.367738-0.064257 i$ & $0.367738-0.064257 i$ & $0\%$\\
\end{tabular}
\end{ruledtabular}
\endgroup
\end{center}

\begin{center}
\inlinetablecaption{Electromagnetic overtones of the $\ell=2$ mode for the Jusufi-Singleton black hole with $M=1$.  The rows are grouped by the overtone number $n=1,2,3$.  The two central columns show the WKB16--Pad\'e and WKB14--Pad\'e estimates, and $\Delta$ is their relative difference in percent.}
\begingroup
\scriptsize
\begin{ruledtabular}
\begin{tabular}{c c c c c}
Mode & $l_0$ & WKB16 ($\tilde{m}=8$) & WKB14 ($\tilde{m}=7$) & $\Delta$ \\
$(2,1)$ & $0.05$ & $0.063612-0.042237 i$ & $0.063612-0.042237 i$ & $0\%$\\
$(2,1)$ & $0.1$ & $0.112056-0.073934 i$ & $0.112056-0.073934 i$ & $0\%$\\
$(2,1)$ & $0.15$ & $0.150730-0.098665 i$ & $0.150730-0.098665 i$ & $0\%$\\
$(2,1)$ & $0.2$ & $0.182674-0.118510 i$ & $0.182674-0.118510 i$ & $0\%$\\
$(2,1)$ & $0.25$ & $0.209795-0.134761 i$ & $0.209795-0.134761 i$ & $0\%$\\
$(2,1)$ & $0.3$ & $0.233382-0.148264 i$ & $0.233382-0.148264 i$ & $0\%$\\
$(2,1)$ & $0.35$ & $0.254360-0.159585 i$ & $0.254360-0.159585 i$ & $0\%$\\
$(2,1)$ & $0.4$ & $0.273434-0.169110 i$ & $0.273434-0.169110 i$ & $0\%$\\
$(2,1)$ & $0.45$ & $0.291163-0.177094 i$ & $0.291163-0.177094 i$ & $0\%$\\
$(2,1)$ & $0.5$ & $0.308020-0.183692 i$ & $0.308020-0.183692 i$ & $0\%$\\
$(2,1)$ & $0.55$ & $0.324426-0.188966 i$ & $0.324426-0.188966 i$ & $0\%$\\
$(2,1)$ & $0.6$ & $0.340781-0.192875 i$ & $0.340781-0.192875 i$ & $0\%$\\
$(2,1)$ & $0.65$ & $0.357498-0.195244 i$ & $0.357498-0.195244 i$ & $0\%$\\
$(2,2)$ & $0.05$ & $0.058502-0.072861 i$ & $0.058501-0.072861 i$ & $0.0002\%$\\
$(2,2)$ & $0.1$ & $0.103207-0.127485 i$ & $0.103207-0.127485 i$ & $0.0002\%$\\
$(2,2)$ & $0.15$ & $0.139081-0.170039 i$ & $0.139081-0.170039 i$ & $0.0002\%$\\
$(2,2)$ & $0.2$ & $0.168898-0.204119 i$ & $0.168898-0.204119 i$ & $0.0002\%$\\
$(2,2)$ & $0.25$ & $0.194397-0.231960 i$ & $0.194397-0.231960 i$ & $0.0002\%$\\
$(2,2)$ & $0.3$ & $0.216761-0.255020 i$ & $0.216761-0.255020 i$ & $0\%$\\
$(2,2)$ & $0.35$ & $0.236849-0.274275 i$ & $0.236848-0.274274 i$ & $0.0002\%$\\
$(2,2)$ & $0.4$ & $0.255321-0.290384 i$ & $0.255321-0.290383 i$ & $0.0002\%$\\
$(2,2)$ & $0.45$ & $0.272709-0.303779 i$ & $0.272710-0.303779 i$ & $0\%$\\
$(2,2)$ & $0.5$ & $0.289466-0.314715 i$ & $0.289465-0.314715 i$ & $0.0002\%$\\
$(2,2)$ & $0.55$ & $0.305993-0.323284 i$ & $0.305992-0.323284 i$ & $0.0003\%$\\
$(2,2)$ & $0.6$ & $0.322664-0.329399 i$ & $0.322663-0.329399 i$ & $0.0003\%$\\
$(2,2)$ & $0.65$ & $0.339829-0.332727 i$ & $0.339828-0.332726 i$ & $0.00028\%$\\
$(2,3)$ & $0.05$ & $0.052920-0.106047 i$ & $0.052920-0.106048 i$ & $0.0013\%$\\
$(2,3)$ & $0.1$ & $0.093530-0.185467 i$ & $0.093530-0.185469 i$ & $0.0013\%$\\
$(2,3)$ & $0.15$ & $0.126322-0.247239 i$ & $0.126322-0.247243 i$ & $0.0015\%$\\
$(2,3)$ & $0.2$ & $0.153781-0.296606 i$ & $0.153780-0.296611 i$ & $0.0016\%$\\
$(2,3)$ & $0.25$ & $0.177465-0.336833 i$ & $0.177462-0.336839 i$ & $0.0015\%$\\
$(2,3)$ & $0.3$ & $0.198439-0.370042 i$ & $0.198438-0.370045 i$ & $0.0008\%$\\
$(2,3)$ & $0.35$ & $0.217479-0.397644 i$ & $0.217483-0.397647 i$ & $0.0011\%$\\
$(2,3)$ & $0.4$ & $0.235233-0.420607 i$ & $0.235218-0.420605 i$ & $0.0032\%$\\
$(2,3)$ & $0.45$ & $0.252139-0.439524 i$ & $0.252134-0.439526 i$ & $0.00101\%$\\
$(2,3)$ & $0.5$ & $0.268650-0.454768 i$ & $0.268650-0.454768 i$ & $0\%$\\
$(2,3)$ & $0.55$ & $0.285130-0.466442 i$ & $0.285129-0.466444 i$ & $0.00040\%$\\
$(2,3)$ & $0.6$ & $0.301885-0.474389 i$ & $0.301883-0.474395 i$ & $0.00112\%$\\
$(2,3)$ & $0.65$ & $0.319135-0.478086 i$ & $0.319131-0.478094 i$ & $0.00152\%$\\
\end{tabular}
\end{ruledtabular}
\endgroup
\end{center}

\begin{center}
\inlinetablecaption{Fundamental Dirac quasinormal modes of the Jusufi-Singleton black hole for $M=1$.  The two isospectral Dirac potentials give the same spectrum, so the listed modes are labeled only by $(\ell,n)$.  The last column gives the relative difference between the two WKB--Pad\'e estimates in percent.}
\begingroup
\scriptsize
\begin{ruledtabular}
\begin{tabular}{c c c c c}
Mode & $l_0$ & WKB16 ($\tilde{m}=8$) & WKB14 ($\tilde{m}=7$) & $\Delta$ \\
$(1/2,0)$ & $0.05$ & $0.026627-0.014134 i$ & $0.026639-0.014108 i$ & $0.093\%$\\
$(1/2,0)$ & $0.1$ & $0.046887-0.024734 i$ & $0.046906-0.024695 i$ & $0.0824\%$\\
$(1/2,0)$ & $0.15$ & $0.063040-0.033000 i$ & $0.063059-0.032953 i$ & $0.0714\%$\\
$(1/2,0)$ & $0.2$ & $0.076360-0.039628 i$ & $0.076376-0.039578 i$ & $0.0617\%$\\
$(1/2,0)$ & $0.25$ & $0.087645-0.045052 i$ & $0.087656-0.045000 i$ & $0.0532\%$\\
$(1/2,0)$ & $0.3$ & $0.097434-0.049552 i$ & $0.097438-0.049502 i$ & $0.0456\%$\\
$(1/2,0)$ & $0.35$ & $0.106111-0.053317 i$ & $0.106107-0.053272 i$ & $0.0384\%$\\
$(1/2,0)$ & $0.4$ & $0.113966-0.056476 i$ & $0.113955-0.056437 i$ & $0.0314\%$\\
$(1/2,0)$ & $0.45$ & $0.121226-0.059111 i$ & $0.121210-0.059082 i$ & $0.0249\%$\\
$(1/2,0)$ & $0.5$ & $0.128082-0.061277 i$ & $0.128064-0.061257 i$ & $0.0193\%$\\
$(1/2,0)$ & $0.55$ & $0.134702-0.062995 i$ & $0.134684-0.062982 i$ & $0.0150\%$\\
$(1/2,0)$ & $0.6$ & $0.141245-0.064251 i$ & $0.141228-0.064244 i$ & $0.0120\%$\\
$(1/2,0)$ & $0.65$ & $0.147873-0.064986 i$ & $0.147857-0.064983 i$ & $0.0101\%$\\
$(3/2,0)$ & $0.05$ & $0.055353-0.014007 i$ & $0.055353-0.014007 i$ & $0\%$\\
$(3/2,0)$ & $0.1$ & $0.097414-0.024520 i$ & $0.097414-0.024520 i$ & $0\%$\\
$(3/2,0)$ & $0.15$ & $0.130878-0.032725 i$ & $0.130878-0.032725 i$ & $0\%$\\
$(3/2,0)$ & $0.2$ & $0.158405-0.039311 i$ & $0.158405-0.039311 i$ & $0\%$\\
$(3/2,0)$ & $0.25$ & $0.181662-0.044708 i$ & $0.181662-0.044708 i$ & $0\%$\\
$(3/2,0)$ & $0.3$ & $0.201770-0.049194 i$ & $0.201770-0.049194 i$ & $0\%$\\
$(3/2,0)$ & $0.35$ & $0.219529-0.052960 i$ & $0.219529-0.052960 i$ & $0\%$\\
$(3/2,0)$ & $0.4$ & $0.235541-0.056132 i$ & $0.235541-0.056132 i$ & $0\%$\\
$(3/2,0)$ & $0.45$ & $0.250280-0.058798 i$ & $0.250280-0.058798 i$ & $0\%$\\
$(3/2,0)$ & $0.5$ & $0.264138-0.061009 i$ & $0.264138-0.061009 i$ & $0\%$\\
$(3/2,0)$ & $0.55$ & $0.277464-0.062788 i$ & $0.277464-0.062788 i$ & $0\%$\\
$(3/2,0)$ & $0.6$ & $0.290586-0.064125 i$ & $0.290586-0.064125 i$ & $0\%$\\
$(3/2,0)$ & $0.65$ & $0.303845-0.064964 i$ & $0.303845-0.064964 i$ & $0\%$\\
\end{tabular}
\end{ruledtabular}
\endgroup
\end{center}
%\clearpage

The time-domain profiles used for the comparison with the WKB16--Pad\'e values are shown in Figs.~\ref{fig:tdscalar}--\ref{fig:tddirac}.  All profiles correspond to $l_0=0.65$ and $M=1$; the quoted discrepancies are computed as $\delta=|\omega_{\rm P}-\omega_{\rm WKB}|/|\omega_{\rm WKB}|\times100\%$.

The results show a clear and uniform dependence on the zero-point length $l_0$.  For all three kinds of perturbations the real part of the frequency grows monotonically as $l_0$ is increased.  This behavior is consistent with the potentials in Figs.~\ref{fig:pot1}--\ref{fig:pot3}: a larger value of $l_0$ raises the height of the effective barrier, and therefore the characteristic oscillation frequency of the trapped wave packet also increases.  The effect is visible already for the lowest modes and remains present for higher multipoles.  For example, the scalar fundamental frequencies move from the small values characteristic of the shallow barrier at $l_0=0.05$ to substantially larger oscillation frequencies at $l_0=0.65$.  The same trend is seen in the electromagnetic and Dirac spectra.

The imaginary part is negative in all cases, as expected for stable ringdown modes.  Increasing $l_0$ generally makes the magnitude of $\im{\omega}$ larger, so the perturbations decay faster when the non-local correction is stronger.  For the fundamental modes this increase is most pronounced at small and intermediate $l_0$, while at the upper end of the interval the damping rate changes more slowly and tends to approach values of order $|\im{\omega}|\simeq 0.06$--$0.07$.  The overtones of the electromagnetic $\ell=2$ mode follow the same qualitative dependence on $l_0$, but their damping rates are much larger, as expected from the increasing overtone number.  Thus the new parameter affects both parts of the spectrum in the same qualitative direction: it raises the oscillation frequency and shortens the damping time.\\

\onecolumngrid
\begin{center}
\resizebox{0.48\textwidth}{!}{\includegraphics{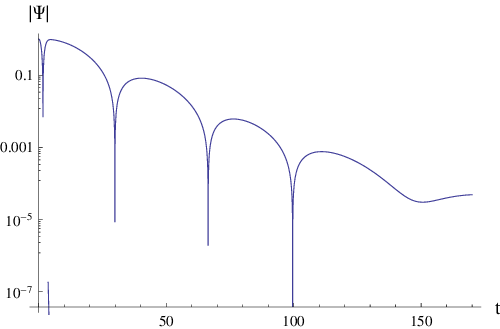}}\hfill
\resizebox{0.48\textwidth}{!}{\includegraphics{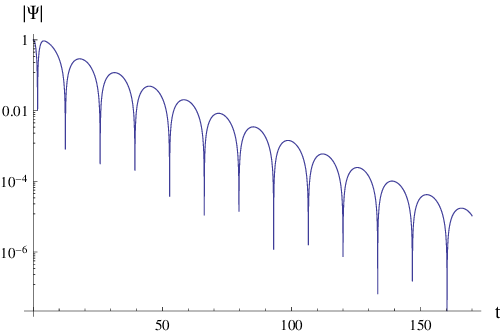}}\\
\inlinefigurecaption{Time-domain profiles for scalar perturbations of the Jusufi-Singleton black hole at $l_0=0.65$ and $M=1$.  The left panel shows the $\ell=0$ mode and the right panel shows the $\ell=1$ mode.  The Prony/WKB16 comparison gives $\ell=0$: $\omega_{\rm P}=0.088238-0.069361 i$ versus $\omega_{\rm WKB}=0.088221-0.069215 i$ ($\delta=0.131\%$), and $\ell=1$: $\omega_{\rm P}=0.233741-0.065664 i$ versus $\omega_{\rm WKB}=0.233741-0.065665 i$ ($\delta=0.000235\%$), where $\delta=|\omega_{\rm P}-\omega_{\rm WKB}|/|\omega_{\rm WKB}|\times100\%$.}{fig:tdscalar}
\end{center}

\begin{center}
\resizebox{0.48\textwidth}{!}{\includegraphics{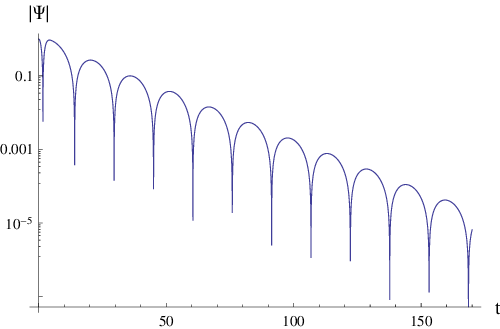}}\hfill
\resizebox{0.48\textwidth}{!}{\includegraphics{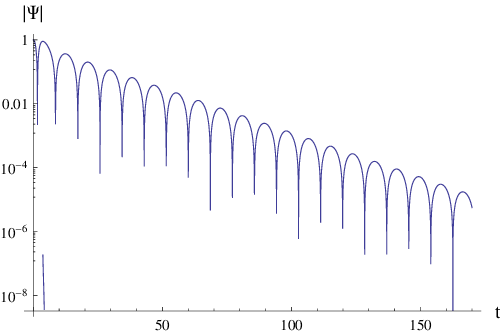}}\\
\inlinefigurecaption{Time-domain profiles for electromagnetic perturbations at $l_0=0.65$ and $M=1$.  The left and right panels correspond to $\ell=1$ and $\ell=2$, respectively.  The Prony/WKB16 comparison gives $\ell=1$: $\omega_{\rm P}=0.203462-0.062956 i$ versus $\omega_{\rm WKB}=0.203462-0.062956 i$ ($\delta=6.58\times10^{-5}\%$), and $\ell=2$: $\omega_{\rm P}=0.367738-0.064257 i$ versus $\omega_{\rm WKB}=0.367738-0.064257 i$ ($\delta=4.38\times10^{-5}\%$).}{fig:tdem}
\end{center}

\begin{center}
\resizebox{0.48\textwidth}{!}{\includegraphics{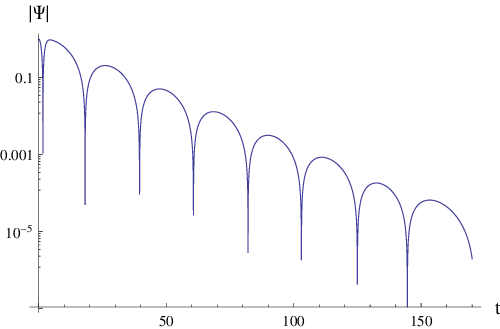}}\hfill
\resizebox{0.48\textwidth}{!}{\includegraphics{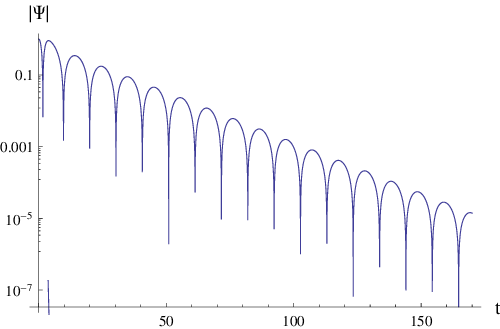}}\\
\inlinefigurecaption{Time-domain profiles for Dirac perturbations at $l_0=0.65$ and $M=1$.  The left panel shows $\ell=1/2$, while the right panel shows $\ell=3/2$.  The Prony/WKB16 comparison gives $\ell=1/2$: $\omega_{\rm P}=0.147757-0.064868 i$ versus $\omega_{\rm WKB}=0.147873-0.064986 i$ ($\delta=0.102\%$), and $\ell=3/2$: $\omega_{\rm P}=0.303848-0.064976 i$ versus $\omega_{\rm WKB}=0.303845-0.064964 i$ ($\delta=0.00390\%$).  Since the two Dirac effective potentials are isospectral, the profiles represent the same quasinormal spectrum for either Darboux-related potential.}{fig:tddirac}
\end{center}
\twocolumngrid

The comparison between the two WKB--Pad\'e orders suggests that the high-order WKB treatment is very accurate for the modes considered here.  The agreement is especially strong for modes with larger multipole number, where the relative difference between the 16th-order and 14th-order estimates is either zero at the displayed precision or far below one percent.  The least favorable cases are the lowest multipoles, in particular the scalar monopole and the Dirac $\ell=1/2$ mode, because WKB methods are known to work best when the angular momentum barrier is sufficiently pronounced.  Even there, however, the quoted differences remain small: the scalar monopole is at the level of a few tenths of a percent, and the Dirac lowest mode is below the percent level throughout the table.

The time-domain integration gives a more independent test because it does not use the expansion around the peak of the potential.  The Prony frequencies extracted from the ringing profiles at $l_0=0.65$ agree very well with the WKB16--Pad\'e values.  The deviations are about $0.131\%$ for the scalar monopole and $0.102\%$ for the lowest Dirac mode, while for the higher scalar, electromagnetic and Dirac multipoles they are much smaller.  This confirms that the WKB results are reliable for the main physical conclusions drawn here.  The only caution is the standard one: the lowest multipoles are the most sensitive to the approximation scheme, so they should be interpreted with slightly larger numerical uncertainty than the higher-$\ell$ modes.

\begin{center}
\resizebox{\linewidth}{!}{\includegraphics{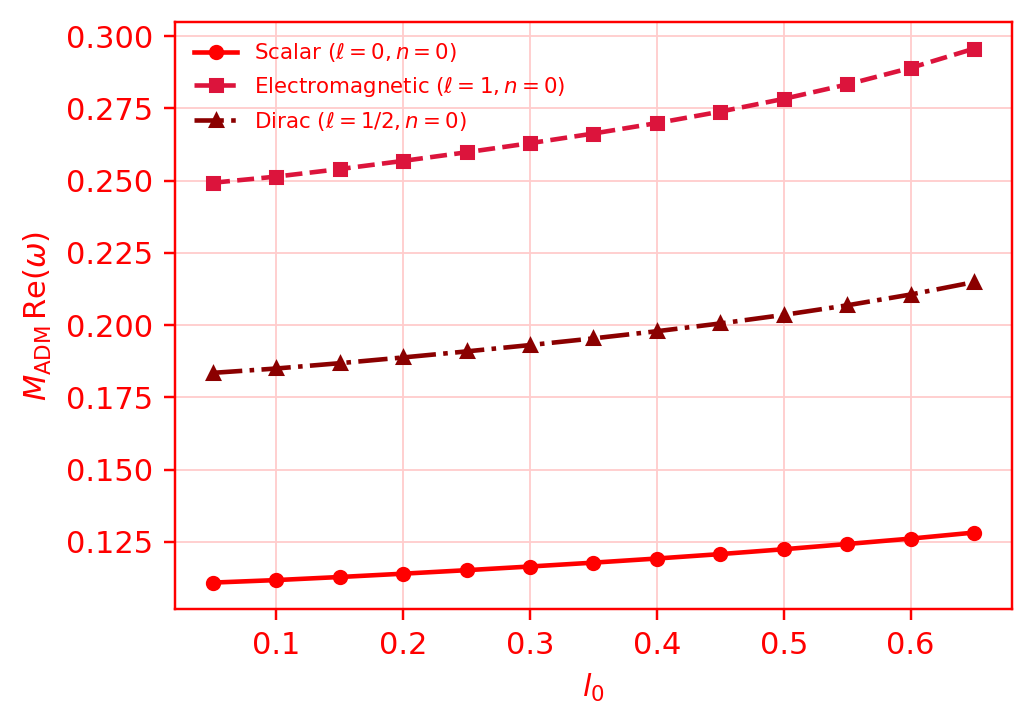}}
{\inlinefigurecaption{ADM-rescaled real part of the representative fundamental quasinormal frequencies.  The plotted quantity is $M_{\rm ADM}\,\mathrm{Re}(\omega)$, obtained directly from the WKB16 data listed above by using $M_{\rm ADM}=1+3\pi/(32l_0)$.}{fig:qnm-adm-real}}
\end{center}

\begin{center}
\resizebox{\linewidth}{!}{\includegraphics{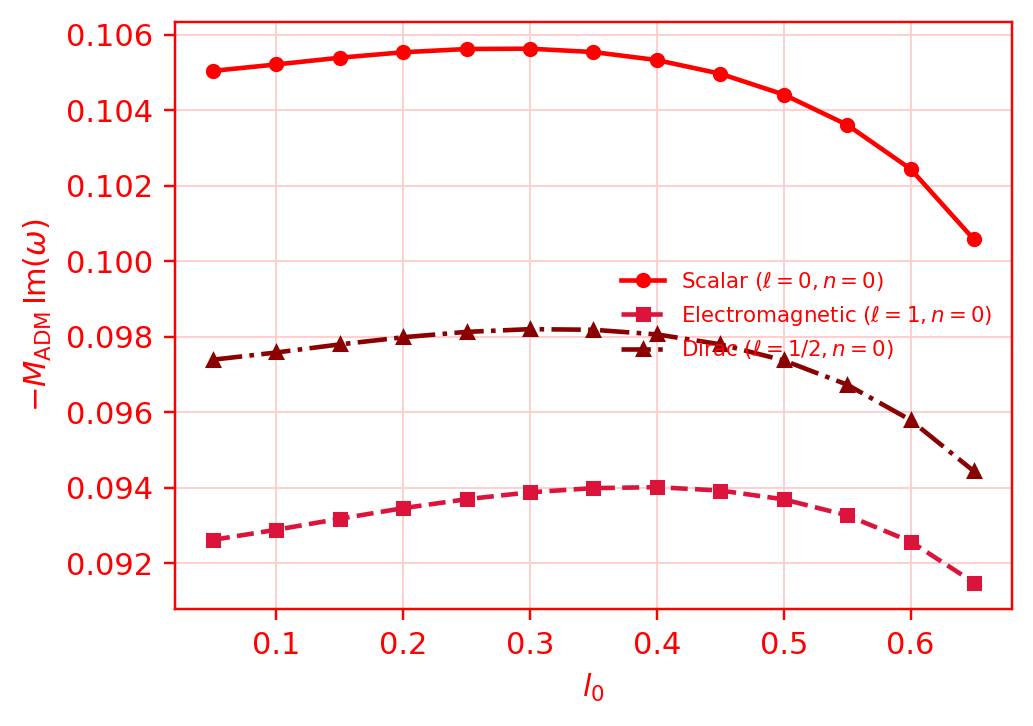}}
{\inlinefigurecaption{ADM-rescaled damping rates for the same representative fundamental modes.  The plotted quantity is $-M_{\rm ADM}\,\mathrm{Im}(\omega)$, so positive values correspond to decaying modes.}{fig:qnm-adm-damping}}
\end{center}

Since the parameter $M$ used in the numerical tables is not the ADM mass, the ADM-normalized frequencies are obtained by multiplying every tabulated $M=1$ frequency by
\[
M_{\rm ADM}=1+\frac{3\pi}{32l_0},
\qquad \bar{\omega}_{\rm ADM}=M_{\rm ADM}\omega .
\]
For the plots below (see figs. \ref{fig:qnm-adm-real} and \ref{fig:qnm-adm-damping}) we use the WKB16 entries for the fundamental scalar, electromagnetic and Dirac modes and display the real oscillation frequency and the damping rate in ADM units.

\section{Excitation factors}\label{sec:excitation}

The same Green-function residue technique used for the Schwarzschild problem can also be applied to the present regular black hole.  The reason is that the perturbation equations have the same one-dimensional scattering form as in Eq.~(\ref{wave-equation}), with a regular event horizon and an asymptotically flat exterior.  For a horizon-normalized solution $\Psi^{\rm in}$ we impose
\begin{equation}
\Psi^{\rm in}\sim e^{-\imo\omega r_*},\qquad r\to r_+,
\end{equation}
where $r_+$ is the largest root of $f(r)=0$.  At infinity the same solution is written as
\begin{equation}\label{Apmdef}
\Psi^{\rm in}\sim A^{(-)}(\omega)e^{-\imo\omega r_*}
+A^{(+)}(\omega)e^{+\imo\omega r_*},\qquad r\to\infty .
\end{equation}
The quasinormal frequencies are therefore the zeros of the incoming amplitude,
\begin{equation}
A^{(-)}(\omega_{\ell n})=0.
\end{equation}
With this normalization the excitation factor is the residue of the frequency-domain Green function at the corresponding pole \cite{Leaver:1986gd,Andersson:1995zk,Berti:2006wq,Zhang:2013ksa},
\begin{equation}\label{Bfactor}
B_{\ell n}=\left.
\frac{A^{(+)}(\omega)}{2\omega\,dA^{(-)}(\omega)/d\omega}
\right|_{\omega=\omega_{\ell n}} .
\end{equation}
It is important to stress that $B_{\ell n}$ is source independent.  A physical ringdown amplitude is obtained only after multiplying it by the appropriate source integral or initial-data coefficient.

For the present metric the tortoise coordinate is fixed by $dr_*/dr=1/f(r)$.  Since excitation factors are complex residues, their phase depends on the additive convention for $r_*$.  In the numerical calculation we use the asymptotic normalization
\begin{equation}\label{rstarconvention}
 r_* - r -2M_{\rm ADM}\ln r \longrightarrow 0,\qquad r\to\infty .
\end{equation}
For $M=1$ the large-distance expansion of the metric function is
\begin{equation}\label{ADMmass}
\begin{split}
 f(r)&=1-\frac{2M_{\rm ADM}}{r}
 +\frac{1+3l_0^2/8}{r^2}+\Order{r^{-3}},\\
 M_{\rm ADM}&=1+\frac{3\pi}{32l_0}.
\end{split}
\end{equation}
Thus, when $l_0$ is small and frequencies are measured in units of $M_{\rm ADM}$, the exterior geometry tends to the Schwarzschild problem, while the additional $1/r^2$ and higher-order terms become subleading.  This provides a useful check on the calculation: the entries with $l_0=0.05$ have $M_{\rm ADM}\omega=0.11098-0.10481\imo$ for the scalar monopole, $0.24927-0.09262\imo$ for the electromagnetic dipole and $0.18368-0.09710\imo$ for the lowest Dirac mode, in agreement with the expected Schwarzschild scale.

With the same asymptotic phase convention, the excitation factors also reproduce the corresponding Schwarzschild residues in the $l_0\to0$ limit.  Numerically, the first rows of Table~V are already close to this limit: after the ADM rescaling the frequencies approach the Schwarzschild values, and the residues tend smoothly to the Schwarzschild excitation factors for the scalar, electromagnetic and Dirac test fields.  For tabulated benchmark data in the standard Schwarzschild/Kerr normalization, see the classic Schwarzschild calculations and the public Kerr excitation-factor data set, whose $a/M=0$ entries give the Schwarzschild values for the scalar and electromagnetic sectors~\cite{Leaver:1986gd,Andersson:1995zk,Berti:2006wq}; the Dirac limit was checked with the same finite-radius matching normalization used here.

Because $B_{\ell n}$ is a complex residue, both its real and imaginary parts may have either sign.  Thus a value such as $-0.01385-0.08146\imo$ is allowed: it simply means that, in the phase convention used here, the residue lies in the third quadrant of the complex plane.  Changing the additive constant in $r_*$, or equivalently rephasing the ingoing/outgoing basis, would rotate the residue by an overall phase, so the sign pattern should not be interpreted as an instability or as a separate physical damping sign.

The numerical procedure is as follows.  We write
\begin{equation}
\Psi^{\rm in}=e^{-\imo\omega r_*}H(r),
\end{equation}
so that the function $H(r)$ is regular at the event horizon.  It obeys
\begin{equation}\label{Heq}
 f^2H''+f\left(f'-2\imo\omega\right)H'-V(r)H=0,
\end{equation}
where primes denote derivatives with respect to $r$.  The integration is started at $r=r_++\epsilon$ with $\epsilon=10^{-5}$, using the regular near-horizon expansion of Eq.~(\ref{Heq}).  For the Dirac case the same procedure is applied to the $V_+$ potential; the Darboux-related $V_-$ potential gives the same QNM frequencies, but the phase convention for the wave function would change the quoted residue.  At a finite radius $R$ the numerical solution is matched to the two asymptotic series
\begin{eqnarray}
\Psi^{\rm in}&=&A^{(-)}e^{-\imo\omega r_*}
\sum_{k=0}^{K}\frac{a_k^{(-)}}{r^k}\nonumber\\
&&+A^{(+)}e^{+\imo\omega r_*}
\sum_{k=0}^{K}\frac{a_k^{(+)}}{r^k},
\qquad a_0^{(\pm)}=1 .
\end{eqnarray}
The coefficients $a_k^{(\pm)}$ are obtained recursively by substituting the series into Eq.~(\ref{wave-equation}) together with the large-$r$ expansion of $f(r)$ and of the relevant potential.  We use $K=12$ and a five-point finite-difference stencil with step $10^{-5}$ for $dA^{(-)}/d\omega$.  The matching radii are chosen as $R=400,220,120,100,100$ for $l_0=0.05,0.1,0.3,0.5,0.65$, respectively.  Varying $K$ and $R$ shows that the lowest scalar mode is the most sensitive case, while the electromagnetic and Dirac entries are considerably more stable.

\onecolumngrid
\begin{center}
\inlinetablecaption{Excitation factors for representative fundamental modes of the Jusufi-Singleton black hole.  The frequencies are the roots of $A^{(-)}(\omega)=0$ obtained by finite-radius matching and are shown in units $M=1$.  The excitation factor $B_{\ell n}$ is defined by Eq.~(\ref{Bfactor}) with the tortoise-coordinate convention of Eq.~(\ref{rstarconvention}).}
\begingroup
\scriptsize
\begin{ruledtabular}
\begin{tabular}{c c c c c}
Mode & $l_0$ & $\omega$ & $M_{\rm ADM}\omega$ & $B_{\ell 0}$ \\
Scalar $(\ell=0)$ & $0.05$ & $0.016107-0.015211\imo$ & $0.11098-0.10481\imo$ & $0.00935-0.07145\imo$\\
Scalar $(\ell=0)$ & $0.1$ & $0.028369-0.026629\imo$ & $0.11192-0.10506\imo$ & $0.03396-0.08450\imo$\\
Scalar $(\ell=0)$ & $0.3$ & $0.058701-0.053202\imo$ & $0.11633-0.10543\imo$ & $0.07718-0.09788\imo$\\
Scalar $(\ell=0)$ & $0.5$ & $0.076933-0.065666\imo$ & $0.12225-0.10435\imo$ & $0.09915-0.09878\imo$\\
Scalar $(\ell=0)$ & $0.65$ & $0.088230-0.069373\imo$ & $0.12821-0.10081\imo$ & $0.11058-0.09374\imo$\\
Electromagnetic $(\ell=1)$ & $0.05$ & $0.036176-0.013441\imo$ & $0.24927-0.09262\imo$ & $0.05500+0.02705\imo$\\
Electromagnetic $(\ell=1)$ & $0.1$ & $0.063722-0.023543\imo$ & $0.25140-0.09288\imo$ & $0.04037+0.06358\imo$\\
Electromagnetic $(\ell=1)$ & $0.3$ & $0.132687-0.047370\imo$ & $0.26295-0.09388\imo$ & $-0.01191+0.09675\imo$\\
Electromagnetic $(\ell=1)$ & $0.5$ & $0.175130-0.058959\imo$ & $0.27829-0.09369\imo$ & $-0.04018+0.09835\imo$\\
Electromagnetic $(\ell=1)$ & $0.65$ & $0.203462-0.062956\imo$ & $0.29565-0.09148\imo$ & $-0.05824+0.09242\imo$\\
Dirac $(\ell=1/2)$ & $0.05$ & $0.026656-0.014091\imo$ & $0.18368-0.09710\imo$ & $-0.03599-0.05591\imo$\\
Dirac $(\ell=1/2)$ & $0.1$ & $0.046934-0.024667\imo$ & $0.18516-0.09732\imo$ & $-0.01385-0.08146\imo$\\
Dirac $(\ell=1/2)$ & $0.3$ & $0.097471-0.049452\imo$ & $0.19316-0.09800\imo$ & $0.03581-0.10269\imo$\\
Dirac $(\ell=1/2)$ & $0.5$ & $0.128086-0.061210\imo$ & $0.20353-0.09727\imo$ & $0.06014-0.10335\imo$\\
Dirac $(\ell=1/2)$ & $0.65$ & $0.147881-0.064917\imo$ & $0.21489-0.09433\imo$ & $0.07445-0.09826\imo$\\
\end{tabular}
\end{ruledtabular}
\endgroup
\end{center}
\twocolumngrid

The excitation factors vary smoothly with the non-local parameter.  For the scalar monopole the magnitude of $B_{00}$ grows as $l_0$ is increased, indicating that the residue of the fundamental pole becomes larger as the effective barrier is raised.  The electromagnetic and Dirac sectors show a more pronounced phase rotation in the complex plane, while their absolute values remain of the same order across the interval considered here.  This behavior should not be confused with the total waveform amplitude: the latter also depends on the source integral.  The table therefore characterizes the intrinsic pole strength of the background, not the response to a particular initial perturbation.

\section{Conclusions}

The quasinormal spectra show that the zero-point length $l_0$ leaves a robust and systematic imprint on the ringdown of the Jusufi-Singleton black hole.  For scalar, electromagnetic and Dirac perturbations the real part of the fundamental frequency increases as $l_0$ grows.  This trend follows the growth of the effective-potential barrier and is therefore common to all three spin sectors.  The imaginary part remains negative throughout the parameter range, and its magnitude generally increases with $l_0$, which means that stronger non-local corrections lead to faster damping.  The growth of the damping rate is most visible at small and intermediate $l_0$, while for the largest values considered here it changes more slowly.

The ordering of the spectra is also physically transparent.  The scalar monopole has the lowest oscillation frequency, whereas the electromagnetic dipole and the Dirac lowest mode are shifted upward by their stronger angular barrier.  Higher electromagnetic overtones have substantially larger damping rates, as expected, but they preserve the same qualitative dependence on the non-local parameter.  Thus the main effect of $l_0$ is not to introduce qualitatively new branches in the range studied here, but to move the usual black-hole ringing frequencies in a controlled way.

The numerical checks support the reliability of these conclusions.  The comparison between the 16th-order and 14th-order WKB--Pad\'e estimates gives very small relative differences for most modes, often vanishing at the displayed precision for the higher multipoles.  The largest uncertainty is confined to the lowest barriers, especially the scalar monopole and the lowest Dirac mode, where WKB methods are naturally less optimal; even there the discrepancy stays at the level of a few tenths of a percent or below.  The time-domain integration provides a more independent validation.  At $l_0=0.65$ the Prony frequencies agree with the WKB16--Pad\'e values at the $0.131\%$ level for the scalar monopole and at the $0.102\%$ level for the lowest Dirac mode, with still smaller differences for the higher multipoles.

We have also evaluated the source-independent excitation factors associated with the fundamental poles.  The residue calculation can be carried over from the Schwarzschild case once the same scattering normalization is fixed; in this work the phase was fixed by the asymptotic convention $r_*-r-2M_{\rm ADM}\ln r\to0$.  In the small-$l_0$ regime the ADM-rescaled frequencies approach the expected Schwarzschild scale, which gives a useful check on the normalization.  The excitation factors vary smoothly with $l_0$: the scalar monopole residue grows in magnitude, while the electromagnetic and Dirac residues mainly rotate in the complex plane and remain of comparable size.  These factors should be regarded as intrinsic pole strengths, not complete waveform amplitudes, because the latter still depend on the chosen source or initial data.

Using the quasinormal frequencies obtained here together with the recently proposed correspondence between quasinormal modes and grey-body factors  \cite{Konoplya:2024lir,Malik:2024cgb,Skvortsova:2024msa,Bolokhov:2024otn,Lutfuoglu:2025mqa,Han:2025cal,Han:2026fpn,Dubinsky:2024vbn,Konoplya:2024vuj}, one can estimate the corresponding transmission probabilities without solving the scattering problem directly. This could provide an additional consistency check of the correspondence in the case of black holes endowed with primary scalar hair.

Overall, the zero-point-length correction produces a clear spectral signature: larger $l_0$ increases the oscillation frequencies, tends to shorten the damping time, and modifies the pole residues without spoiling the single-barrier black-hole ringing picture.  A natural extension would be to compute excitation coefficients for explicit sources or initial data, where the excitation factors obtained here would enter the actual ringdown amplitudes.

\begin{acknowledgments}
The author would like to thank Sergei Bolokhov for useful discussions. 
\end{acknowledgments}

\bibliography{bibliography}
\end{document}